\documentclass[tex,twocolumn,epjc3,tightenlines]{svjour3}
\RequirePackage[T1]{fontenc}
\RequirePackage{graphicx}
\RequirePackage{mathptmx}      
\RequirePackage[numbers,sort&compress]{natbib}
\RequirePackage[colorlinks,citecolor=blue,urlcolor=blue,linkcolor=blue]{hyperref}
\usepackage{amsmath, amssymb}
\usepackage{graphics,bm}
\usepackage{graphicx}
\usepackage{bbold}
\usepackage{slashed}
\usepackage{feynmf}
\usepackage{physics}
\usepackage{wrapfig}
\usepackage{hyperref}

\usepackage{tikz}
\usetikzlibrary{arrows,shapes}
\usetikzlibrary{trees}
\usetikzlibrary{matrix,arrows} 			
\usetikzlibrary{positioning}				
\usetikzlibrary{calc,through}				
\usepackage{pgffor}							

\usetikzlibrary{decorations.pathmorphing}	
\usetikzlibrary{decorations.markings}
\tikzset{
    sigmaCT/.style={draw=black, postaction={decorate},
        decoration={markings,mark=at position .99 with {\arrow[draw=black]{>}},mark=at 		 position .99 with {\arrow[draw=black]{<}}}},
    pionCT/.style={dashed,draw=black, postaction={decorate},
        decoration={markings,mark=at position .99 with {\arrow[draw=black]{>}},mark=at position .99 with {\arrow[draw=black]{<}}}},    
    fermionCT/.style={draw=black, postaction={decorate},
        decoration={markings,mark=at position .5 with {\arrow[draw=black]{>}},mark=at position .99 with {\arrow[draw=black]{>}},mark=at position .99 with {\arrow[draw=black]{<}}}},    
    fermion/.style={draw=black, postaction={decorate},
        decoration={markings,mark=at position .55 with {\arrow[draw=black]{>}}}},
    fermionbar/.style={draw=black, postaction={decorate},
        decoration={markings,mark=at position .55 with {\arrow[draw=black]{<}}}},
    pion/.style={dashed,draw=black, postaction={decorate}},
    sigma/.style={draw=black, postaction={decorate}}
}

\newcommand{\be}{\begin{equation}}
\newcommand{\ee}{\end{equation}}

\newcommand{\bqa}{\begin{eqnarray}}
\newcommand{\eqa}{\end{eqnarray}}

\def\square{\vcenter{\vbox{\hrule height.4pt
          \hbox{\vrule width.4pt height4pt
          \kern4pt\vrule width.3pt}\hrule height.4pt}}}

\journalname{Eur. Phys. J.}

\begin{document}
\title{Two-flavor chiral perturbation theory at nonzero isospin:
Pion condensation at zero temperature}
\author{Prabal Adhikari\thanksref{e1,addr0,addr1,addr2} \and Jens O. Andersen
  \thanksref{e2,addr2}\and Patrick Kneschke\thanksref{e3,addr3}}
\thankstext{e1}{e-mail: pa100@wellesley.edu }

\thankstext{e2}{e-mail: andersen@tf.phys.ntnu.no} 
\thankstext{e3}{e-mail: patrick.kneschke@uis.no}   
\institute{St. Olaf College, Physics Department, 1520 St. Olaf Avenue,
  Northfield, MN 55057, United States 
  \label{addr0}
\and
  Wellesley College, Department of Physics, 106 Central Street,
  Wellesley, MA 02481, United States 
  \label{addr1}
          \and
  Department of Physics, 
Norwegian University of Science and Technology, H{\o}gskoleringen 5,
N-7491 Trondheim, Norway
          \label{addr2}
          \and
Faculty of Science and Technology, University of Stavanger,
N-4036 Stavanger, Norway
          \label{addr3}
}
%
\date{Received: date / Accepted: date}
\date{\today}
\maketitle
\begin{abstract}
  In this paper, we calculate the equation of state of
  two-flavor finite isospin chiral
  perturbation theory at next-to-leading order in the pion-condensed
  phase at zero temperature. We show that the transition from
  the vacuum phase to a Bose-condensed phase is of second order.
  While the tree-level result has been
  known for some time, surprisingly quantum effects have not yet been
  incorporated into the equation of state.
  We find that the corrections to the quantities we compute, 
  namely the isospin density, pressure, and equation of state,
  increase with increasing isospin
  chemical potential. 
  We compare our results to
  recent lattice simulations of 2+1 flavor QCD with physical quark masses.
  The agreement with the lattice results is generally good and improves
  somewhat as we go from leading order to next-to-leading order
  in $\chi$PT. 

\end{abstract}

\section{Introduction}
Quantum chromodynamics (QCD), the fundamental theory of strong interactions,
has a rich phase structure, particularly at finite baryon densities relevant
for a number of physical systems including neutron stars, neutron matter and
heavy-ion
collisions among others~\cite{raja,alford,fukurev}.
However, finite baryon densities are not accessible
directly through QCD since the physics is non-perturbative and lattice
calculations are hindered by the fermion sign problem. Though it is worth
noting that some progress has been made in circumventing the sign problem
through the fermion bag and Lefschetz thimble
approaches~\cite{bed1}.
There is also the
additional possibility of solving QCD at finite baryon density with quantum
computers since the sign problem is absent in quantum
algorithms~\cite{q1}.

While finite baryon density is inaccessible through lattice QCD, finite isospin
systems in real QCD can be studied using lattice-based methods, see
Ref.~\cite{kogut1,kogut2} for some early results.
The most thorough of these studies were performed only
recently~\cite{gergy1,gergy2,gergy3}
even though finite
isospin QCD was first studied 
over a decade ago using chiral perturbation theory ($\chi$PT)
in a seminal paper by Son and
Stephanov~\cite{son}.
$\chi$PT~\cite{wein,gasser1,gasser2,bein}
is a low-energy effective field theory of QCD that describes the dynamics
of the pseudo-Goldstone bosons that are the result of the spontaneous
symmetry breaking of global symmetries in the QCD vacuum.
Being based only on symmetries and degrees of freedom, the predictions
of 
are model independent. 

It is agreed through both lattice QCD and chiral perturbation theory
studies that at an
isospin chemical potential equal to the physical pion mass there
is a second-order phase transition at zero temperature
from the vacuum phase to
a pion-condensed
phase. With increasing chemical potentials there is a crossover transition to a
BCS phase with a parity breaking order parameter,
$\langle\bar{u}\gamma_{5}d\rangle\neq 0$ or
$\langle\bar{d}\gamma_{5}u\rangle\neq 0$, that has the same quantum numbers as
a charged pion condensate.
Furthermore, for large temperatures of
approximately $170$\ MeV, the pion condensate is destroyed due to
thermal fluctuations.
Various aspects of $\chi$PT at finite isospin density can be found in
Refs.~\cite{son,kim,loewe,fragaiso,cohen2,janssen,carig0,carigchpt,luca}.
Finite isospin systems have also been studied in the
context of QCD models including the non-renormalizable
Nambu-Jona-Lasinio model~\cite{2fbuballa,toublannjl,bar2f,he2f,heman2,heman,ebert1,ebert2,sun,lars,2fabuki,heman3,he3f,ricardo,ruggi}, 
and the
renormalizable quark-meson model~\cite{lorenz,ueda,qmstiele,allofus},
with the results found there being largely
in agreement with lattice QCD. A very recent review of
meson condensation can be found in Ref.~\cite{mannarev}.

In addition to the study of pions at finite isospin chemical potential there
has also been recent interest in the study of pions in the presence of 
external magnetic fields, which are relevant in the context of neutron stars
with large fields (magnetars) and possibly in RHIC collisions, which generate
magnetic fields due to accelerated charged beams of lead and gold nuclei. In
neutron star cores, an isospin asymmetry is present since protons are converted
into neutrons and neutrinos through electron capture. However, in the presence
of a magnetic field, finite isospin systems are difficult to study due to the
fermion sign problem on the lattice QCD that arises as a consequence of flavor
asymmetry between up and anti-down quarks for electromagnetic interactions.
The complex action problem is tackled by studying finite isospin densities for
small magnetic fields, where the sign problem is mild.
The lattice observes a diamagnetic phase~\cite{endromag},
while studies in chiral perturbation theory valid for magnetic fields
$eB\ll(4\pi f_{\pi})^{2}$ suggests that pions behave as a type-II
superconductor~\cite{prabal}.

More recently, due to the accessibility of the equation of state (EoS) of
pion degrees of freedom through lattice QCD, there has been a lot of
interest in the possibility of pion stars~\cite{carigchpt,endro},
a type of boson star that does not
require the hypothesized axion, which was initially proposed as a solution to
the strong CP problem in QCD. Pion stars, on the other hand, only require input
from QCD and it is conjectured that pion condensation takes place
in a gas of dense
neutrinos~\cite{tomas}.
Recent work shows that pion stars are typically much larger in size
than neutron stars due to a softer equation of state and that the isospin
chemical potentials at the center of such stars can be as high as $~250$ MeV
for purely pionic stars and smaller for pion stars electromagnetically
neutralized by leptons~\cite{endro}.

The goal of this paper is to revisit the equation of state for finite isospin
QCD in the regime of validity of $\chi$PT, where we expect
$\mu_{I}\ll 4\pi f_{\pi}$. The equation of state (at tree level) was originally
calculated in Ref.~\cite{son}
of QCD. In this paper, we calculate the equation of state within $\chi$PT and
incorporate leading order quantum corrections.

We begin in Sec. 2
with a brief overview of chiral perturbation theory and discuss how to
parametrize the fluctuations around the ground state.
We derive  the Lagrangian that is needed for all
next-to-leading order (NLO) calculations within
$\chi$PT at finite isospin chemical potential allowing for a charged pion
condensate. In Sec. 3, we use this NLO Lagrangian to calculate the 
renormalized one-loop free energy at finite $\mu_I$.
In Sec. 4, we calculate the isospin density, the pressure, and the 
equation of state in the pion-condensed phase. Our results are compared
to those of recent lattice simulations. We summarize our findings in Sec. 5
and present some calculations' details in Appendices A-E.

\section{$\chi$PT Lagrangian at ${\cal O}(p^4)$}
In this section, we discuss the symmetries of two-flavor QCD
QCD and chiral perturbation
theory as a low-energy approximation to it. The two-flavor
Lagrangian is
\bqa
{\cal L}&=&
\bar{\psi}\left(
i/\!\!\!\!\!\!D-m_q
\right)\psi-{1\over4}F_{\mu\nu}^aF^{\mu\nu a}\;,
\eqa
where $m_q={\rm diag}(m_u,m_d)$ is the mass
matrix, $/\!\!\!\!\!\!D=\gamma^{\mu}(\partial_{\mu}-igA_{\mu}^at^a)$
is the covariant derivative, $t^a$ are the Gell-Mann matrices, and
$F_{\mu\nu}^a$ is the field-strength tensor.

For massless quarks, the global symmetries of QCD are
$SU(2)_L\times SU(2)_R\times U(1)_B$, which is reduced to
$SU(2)_V\times U(1)_B$ for nonzero quark masses
in the isospin limit, i.e. for $m_u=m_d$.
If $m_u\neq m_d$, this is further reduced to
$U(1)_{I_3}\times U(1)_B=U(1)_u\times U(1)_d$. Adding a quark chemical
potential $\mu_q$ for each quark, the symmetry is
$U(1)_{I_3}\times U(1)_B=U(1)_u\times U(1)_d$ irrespective of the
quark mass. In the pion-condensed phase, the $U(1)_{I_3}$
symmetry is broken. In the remainder of the paper, we work in the
isospin limit.

We begin with the chiral perturbation theory Lagrangian in the isospin limit
at $\mathcal{O}(p^{2})$
\bqa
\mathcal{L}_{2}=\frac{f^{2}}{4}{\rm Tr}
  \left [\nabla^{\mu} \Sigma^{\dagger} \nabla_{\mu}\Sigma 
    \right ]
+\frac{f^2 m^{2}}{4} {\rm Tr}
\left [ \Sigma+\Sigma^{\dagger}\right ]\; ,
\label{lag0}
\eqa
where $f$ is the (bare) pion decay constant and $m$ is
the (bare) pion mass.
The relation between the physical pion mass $m_{\pi}$ and $m$, and
between the physical pion decay constant $f_{\pi}$ and $f$ are briefly
discussed in \ref{relations}.
The covariant derivatives at finite isospin are defined as follows
\bqa
\nabla_{\mu} \Sigma&\equiv&
\partial_{\mu}\Sigma-i\left [v_{\mu},\Sigma \right]\\ 
\nabla_{\mu} \Sigma^{\dagger}&=&
\partial_{\mu}\Sigma^{\dagger}-i [v_{\mu},\Sigma^{\dagger} ]
\;,
\eqa
where $v_{\mu}=\delta_{\mu 0}\mu_{I}\frac{\tau_{3}}{2}$ with $\mu_{I}$ denoting the
isospin chemical potential and $\tau_{3}$ the third Pauli matrix.

It is well known that chiral perturbation theory encodes the interactions among
the Goldstone bosons (pions) that arise due to the spontaneous breaking of
chiral symmetry by the QCD vacuum, i.e. 
\begin{equation}
\begin{split}
\Sigma_{ji}\equiv\langle\bar{\psi}_{iR}\psi_{jL} \rangle\neq 0
\end{split}
\end{equation}
Under chiral rotations, i.e. $SU(2)_{L}\times SU(2)_{R}$, 
the left-handed and right-handed fields transform as
\begin{equation}
\begin{split}
\psi_{L}&\rightarrow L \psi_{L}\\
\psi_{R}&\rightarrow R \psi_{R}\ .
\end{split}
\end{equation}
As such $\Sigma$ transforms as
\begin{equation}
\begin{split}
\Sigma\rightarrow L\Sigma R^{\dagger}\ .
\end{split}
\end{equation}

\subsection{Ground State}
We briefly review the ground state of $\chi$PT at finite isospin using the
$\mathcal{O}(p^{2})$ Lagrangian.
The static Hamiltonian is 
\bqa\nonumber
{\cal H}^{\rm static}_2&=&{1\over8}f^2\mu_I^2
\Tr\left[\tau_3\Sigma\tau_3\Sigma^{\dagger}-1\right]
-{1\over4}f^2m^2\Tr\left[\Sigma+\Sigma^{\dagger}\right]\;.
\\ &&\label{stath}
\eqa
The ansatz for a $\mu_I$-dependent rotated ground state can be
parametrized by the angle $\alpha$ as~\cite{son}
\bqa
\Sigma_{\alpha}&=&
e^{i\alpha\hat{\phi}_i\tau_i}
=\cos\alpha
  +i\hat{\phi}_{i}\tau_{i}\sin\alpha\ ,
\eqa
where $\hat{\phi}_{i}\hat{\phi}_{i}=1$.
This requirement guarantees that
$\Sigma_{\alpha}^{\dagger}\Sigma_{\alpha}=\mathbb{1}$.
The static
Hamiltonian at $\mathcal{O}(p^{2})$ then becomes
\begin{equation}
\begin{split}
  \mathcal{H}_{2}^{\rm static}&=-\mathcal{L}_{2}=
  -f^{2}m^{2}\cos\alpha-\frac{1}{2}f^2\mu_{I}^{2}\sin^{2}\alpha(\hat{\phi}_{1}^{2}
    +\hat{\phi}_{2}^{2})\;.
\end{split}
\end{equation}
 The first term in Eq.~(\ref{stath}) favors the vacuum direction since
the trace of the Pauli matrices is zero, while  
the second term  favors directions in
isospin space which anticommute with $\tau_3$, i.e. along
$\tau_1$ and $\tau_2$. 
Thus there is competition between these two terms.
We also note that the ground-state energy is minimized for $\hat{\phi}_3=0$.
Thus $\hat{\phi}_{1}^2+\hat{\phi_{2}^2}=1$ and
neutral pions do not condense. By minimizing the above expression with respect to $\alpha$, we get the
well-known result that charged pion condensation occurs for $\mu_{I}\ge m$ with
$\cos\alpha=\frac{m^{2}}{\mu_{I}^{2}}$. For $\mu_I<m$, $\alpha=0$
and $\Sigma=\mathbb{1}$, i.e. the vacuum solution.

\subsection{Parametrizing Fluctuations}
Since the goal of this paper is to study the equation of state of the pion
condensed phase including quantum corrections, it is natural to expand the
$\chi$PT
Lagrangian around the pion condensed ground state.
The Goldstone manifold as a consequence of chiral symmetry breaking is
$SU(2)_{L}\times SU(2)_{R}/SU(2)_{V}$. As such, we proceed by first
parametrizing the condensed vacuum as follows
\bqa
\Sigma_{\alpha}&=&A_{\alpha}\Sigma_{0}A_{\alpha}\;,\\
A_{\alpha}&=&e^{i\frac{\alpha}{2}
  \left (\hat{\phi}_{1}\tau_{1}+\hat{\phi}_{2}\tau_{2}\right )}
=\cos\mbox{$\alpha\over2$}
+i(\hat{\phi}_1\tau_1+\hat{\phi}_2\tau_2)\sin\mbox{$\alpha\over2$}
\;,
\eqa
where we, for the purposes of this paper, choose $\hat{\phi}_{1}=1$ and
$\hat{\phi}_{2}=0$ without any loss of generality. Note that $\alpha=0$
reproduces the normal vacuum with $\Sigma_{0}=\mathbb{1}$ as required. Then the
fluctuations (which are axial) around this condensed vacuum are parametrized as
\bqa
\Sigma&=L_{\alpha}\Sigma_{\alpha}R_{\alpha}^{\dagger}\;,
\label{sigmas}
\eqa
with 
\bqa
\label{lrot}
L_{\alpha}&=&A_{\alpha}UA^{\dagger}_{\alpha}\;,\\
R_{\alpha}&=&A_{\alpha}^{\dagger}U^{\dagger}A_{\alpha}\;.
\label{rrot}
\eqa
We emphasize that the fluctuations parameterized by $L_{\alpha}$ and $R_{\alpha}$
around the ground state depend on $\alpha$ since the broken generators
(of QCD) need to be rotated appropriately as the condensed vacuum rotates with
the angle $\alpha$~\cite{kim}.~\footnote{Consider e.g.
  a theory with an $SO(3)$ symmetric Lagrangian with the ground state
  picking up a vev say in the $z$-direction. If the vev is rotated to the
  $y$-direction, then the (un)broken generators must be rotated
  accordingly.} We discuss this briefly in \ref{rotgen}.
$U$ is an $SU(2)$ matrix that parameterizes the fluctuations around the ground
state:
\begin{equation}
\begin{split}
U=\exp\left (i\frac{\phi_{a}\tau_{a}}{2f} \right )\;.
\end{split}
\end{equation}
With the parameterizations stated above, we get
\bqa
\label{parametrization}
\Sigma&=&A_{\alpha}(U\Sigma_{0}U)A_{\alpha}\;.
\eqa
As we show later in this paper, this parameterization not only produces the
correct linear terms that vanish at $\mathcal{O}(p^{2})$,
the divergences of one-loop diagrams also cancel using
counterterms from the $\mathcal{O}(p^{4})$ Lagrangian. Furthermore, the
parametrization produces a Lagrangian that is canonical in the fluctuations and
has the correct limit when $\alpha=0$, whereby 
\begin{equation}
\begin{split}
\Sigma=U\Sigma_{0}U=U^{2}=\exp\left (i\frac{\phi_{a}\tau_{a}}{f} \right )\ ,
\end{split}
\end{equation}
as expected.

We would like to emphasize the
importance of using $L_{\alpha}$ and $R_{\alpha}$ instead of
$L=U$ and $R=U^{\dagger}$. If the latter set is used, Eq.~(\ref{sigmas})
is replaced by 
\begin{equation}
\label{sigmawrong}
\Sigma_{\rm wrong}=U\Sigma_{\alpha}U=UA_{\alpha}\Sigma_0A_{\alpha}U\ ,
\end{equation}
and one finds that the kinetic term of the Lagrangian is not properly
normalized.
This is in itself not problematic since the canonical 
normalization can be achieved 
by a field redefinition. This field redefinition changes
the mass and interaction terms of the Lagrangian but only at the
minimum of the LO effective potential do the masses coincide with the
correct expressions, Eqs.~(\ref{m1def})--(\ref{m4def}) below.
Moreover, if one computes the 
one-loop effective potential, it turns out that the counterterms
cancel the divergences only at the classical minimum.
Thus one cannot renormalize
the NLO effective potential away from the LO minimum
and therefore not find the NLO minimum, which
shows that the $\Sigma_{\rm wrong}$ in Eq.~(\ref{sigmawrong}) cannot be correct.

\subsection{Leading-order Lagrangian}
Using the parameterization of Eq.~(\ref{parametrization}) discussed above, we
can write down the Lagrangian in terms of the fields $\phi_{a}$, which
parametrizes the Goldstone manifold
\bqa
\mathcal{L}_{2}&=&
\mathcal{L}_2^{\rm static}
+\mathcal{L}_2^{\rm linear}
+\mathcal{L}_2^{\rm quadratic}
+\cdots\;,
\eqa
where
\bqa
\label{lo1}
\mathcal{L}_2^{\rm static}
&=&f^{2}m^{2}\cos\alpha+
{1\over2}f^2\mu_{I}^2\sin^{2}\alpha
\;,\\ \nonumber
  \mathcal{L}_2^{\rm linear}
  &=&f\left (-m^{2}\sin\alpha
    +\mu_{I}^{2}\cos\alpha\sin\alpha\right )\phi_{1}
\label{lo22}
  \\ 
&&+f\mu_{I}\sin\alpha\partial_{0}\phi_{2}\;,\\  \nonumber
\mathcal{L}_2^{\rm quadratic}
&=&\frac{1}{2}(\partial_{\mu}\phi_{a})
(\partial^{\mu}\phi_{a})+\mu_{I}\cos\alpha(\phi_{1}\partial_{0}\phi_{2}-\phi_{2}
\partial_{0}\phi_{1})\\ \nonumber
&&-\frac{1}{2}\left [(m^{2}\cos\alpha-\mu_{I}^{2}\cos2\alpha)\phi_{1}^{2}
\right.\\ \nonumber
&&+(m^{2}\cos\alpha-\mu_{I}^{2}\cos^{2}\alpha)\phi_{2}^{2}\\
&&\left.+(m^{2}\cos\alpha+\mu_{I}^{2}\sin^{2}\alpha)\phi_{3}^{2}\right]
\label{lo2}
\;.
\eqa
The inverse propagator in the $\phi_{a}$ basis is
\bqa
D^{-1}&=
\begin{pmatrix}
D^{-1}_{12}&0\\
0&P^{2}-m_{3}^{2}
\end{pmatrix}\;,\\
D^{-1}_{12}&=
\begin{pmatrix}
P^{2}-m_{1}^{2}&ip_{0}m_{12}\\
-ip_{0}m_{12}&P^{2}-m_{2}^{2}\\
\end{pmatrix}\;,
\eqa
where $P=(p_0,p)$ is the four-momentum, $P^2=p_0^2-p^2$, and 
the masses are
\bqa
\label{m1def}
m_{1}&=&\sqrt{m^{2}\cos\alpha-\mu_{I}^{2}\cos2\alpha}\;,\\
m_{2}&=&\sqrt{m^{2}\cos\alpha-\mu_{I}^{2}\cos^{2}\alpha}\;,\\
m_{12}&=&2\mu_{I}\cos\alpha\;,\\
m_{3}&=&\sqrt{m^{2}\cos\alpha+\mu_{I}^{2}\sin^{2}\alpha}\;,
\label{m4def}
\eqa
and 
with $D_{12}^{-1}$ representing the inverse propagator for the charged pions.
The dispersion relation can be found using the zeros of the inverse
propagator $D^{-1}$. We find that the energies associated with the three pion
modes are as follows
\bqa\nonumber
E_{\pi^{\pm}}^{2}&=&p^2+{1\over2}\left(m_{1}^{2}+m_{2}^{2}+m_{12}^{2}\right)
\\
&&\pm{1\over2}\sqrt{4p^{2}m_{12}^{2}+(m_{1}^{2}+m_{2}^{2}
  +m_{12}^{2})^2-4m_{1}^{2}m_{2}^{2}}\;,
\label{pipo}
\\
E_{\pi^0}^{2}&=&p^{2}+m_{3}^{2}\;.
\label{pipo2}
\eqa
The full propagator can
then be written in terms of the dispersion relations as follows
\bqa
D&=&
\begin{pmatrix}
D_{12}&0\\
0&(p^{2}-m_{3}^{2})^{-1}\\
\end{pmatrix}\;,\\
D_{12}&=&\frac{1}{(p_{0}^{2}-E_{\pi^+}^{2})(p_{0}^{2}-E_{\pi^-}^{2})}
\begin{pmatrix}
P^{2}-m_{2}^{2}&-ip_{0}m_{12}\\
ip_{0}m_{12}&P^{2}-m_{1}^{2}\\
\end{pmatrix}\;.
\eqa
Expanding the Lagrangian  ${\cal L}_2$
beyond the quadratic terms, we get for terms with
three and four fields
\bqa\nonumber
\mathcal{L}_2^{\rm cubic}&=&\frac{(m^{2}-4\mu_{I}^{2}\cos\alpha)
    \sin\alpha}{6f}\phi_{1}(\phi_{a}\phi_{a})
\\ &&
  -\frac{\mu_{I}\sin\alpha}{f}  \left [\phi_{1}^2\partial_0\phi_2
+    \phi_{3}^2\partial_0\phi_2\right]
\;,
\label{cubic}
\\
  \nonumber
\mathcal{L}_{2}^{\rm quartic}&=&\frac{1}{24f^{2}}(\phi_{a}
\phi_{a})\left
    [(m^{2}\cos\alpha-4\mu_{I}^{2}\cos2\alpha)\phi_{1}^{2}
\right.\\ &&\left. \nonumber
    +(m^{2}\cos\alpha-4\mu_{I}^{2}\cos^{2}\alpha)\phi_{2}^{2}
\right.\\ &&\left. \nonumber
    +(m^{2}
    \cos\alpha+4\mu_{I}^{2}\sin^{2}\alpha)\phi_{3}^{2} \right ]\\
  &&\nonumber
  -\frac{\mu_{I}\cos\alpha}{3f^{2}}
  (\phi_{a}
  \phi_{a})(\phi_{1}\partial_{0}\phi_{2}-\phi_{2}\partial_{0}\phi_{1})
\\ &&
  +\frac{1}{6f^{2}}\left [\phi_{a}\phi_{b}
    \partial^{\mu}
    \phi_{a}    \partial^{\mu}\phi_{b}
    -\phi_{a}\phi_{a}\partial_{\mu}
    \phi_{b}
    \partial^{\mu}\phi_{b} \right ]\;.
\label{lo4}
  \eqa
  The Lagrangian in the normal phase can be recovered simply by setting
  $\alpha=0$. Note in particular that
the cubic terms  vanish, ${\cal L}_2^{\rm cubic}=0$.

\subsection{Next-to-leading order Lagrangian}
In order to perform calculations at NLO, we must consider the terms
in the Lagrangian that contribute at ${\cal O}\left(p^4\right)$.
In the notation of Ref.~\cite{gerber}, the relevant terms
are~\footnote{There are additional operators
  with couplings $l_5$--$l_7$ and $h_2$--$h_3$ which are not relevant for
  the present calculation.}
\bqa
\nonumber
{\cal L}_4&=&  {1\over4}l_1\left({\rm Tr}
\left[D_{\mu}\Sigma^{\dagger}D^{\mu}\Sigma\right]\right)^2
\\ \nonumber&&
+{1\over4}l_2{\rm Tr}\left[D_{\mu}\Sigma^{\dagger}D_{\nu}\Sigma\right]
    {\rm Tr}\left[D^{\mu}\Sigma^{\dagger}D^{\nu}\Sigma\right]
\\ && \nonumber
    +{1\over16}(l_3+l_4)m^4({\rm Tr}[\Sigma+\Sigma^{\dagger}])^2
\\&&
    +{1\over8}l_4m^2{\rm Tr}\left[D_{\mu}\Sigma^{\dagger}D^{\mu}\Sigma\right]
    {\rm Tr}[\Sigma+\Sigma^{\dagger}]
+h_1{\rm Tr}m^4    \;,
\label{lag}
\eqa
where $l_1$--$l_4$  and $h_1$ are bare coupling constants.
The bare and renormalized couplings $l_i^r(\Lambda)$, $h_i^r(\Lambda)$
are related by
\bqa
l_{i}&=&
l_i^r(\Lambda)-{\gamma_i\Lambda^{-2\epsilon}
  \over2(4\pi)^{2}}
\left[
{1\over\epsilon}+1
\right ]\;,
\label{lowl}
\\
h_i&=&h_i^r(\Lambda)-
{\delta_i\Lambda^{-2\epsilon}\over2(4\pi)^{2}}
\left[{1\over\epsilon}+1\right ]\;,
\label{low2}
\eqa
where $\gamma_i$ and $\delta_i$
are coefficients, and
  $\Lambda$ is the renormalization scale in the modified minimal
  subtraction ($\overline{\rm MS}$) scheme (see below).
  The renormalized $l_i^r$s and $h_i^r$s are running couplings
  that satisfy renormalization group equations.
  Since the bare couplings are independent of the renormalization scale
  $\Lambda$, differentiation of Eqs.~(\ref{lowl}) --~(\ref{low2})
  immediately yields 
  \begin{align}
    \label{rgrun}
  \Lambda{d\over d\Lambda}l_i^r&=-{\gamma_i\over(4\pi)^2}\;,
&  \Lambda{d\over d\Lambda}h_i^r=-{\delta_i\over(4\pi)^2}\;.
  \end{align}
  The low-energy constants $\bar{l}_i$ and $\bar{h}_1$ are defined via
the solutions to the renormalization group equations (\ref{rgrun}) as
  \bqa
l_i^r(\Lambda)&=&{\gamma_i\over2(4\pi)^2}\left[\bar{l}_i+\log{m^2\over\Lambda^2}
  \right]\;,\\
h_i^r(\Lambda)&=&{\delta_i\over2(4\pi)^2}\left[\bar{h}_i+\log{m^2\over\Lambda^2}
\right]\;,
\label{hr}
\eqa
and are up to a constant equal to the renormalized couplings
$l_i^r(\Lambda)$  and $h_i^r(\Lambda)$
evaluated at the scale $\Lambda=m$~\cite{gasser1}.
The coefficients $\gamma_i$ and $h_i$ are
\bqa
\gamma_{1}&=&\frac{1}{3}\;,
\hspace{0.3cm}
\gamma_{2}=\frac{2}{3}\;,
\hspace{0.3cm}\gamma_{3}=-\frac{1}{2}\;,
  \\
 \gamma_{4}&=&2\;,  \hspace{0.3cm}\delta_1=0\;.
  \eqa
  Since $\delta_1=0$, Eqs.~(\ref{low2}), (\ref{rgrun}), and (\ref{hr})
  obviously do not apply. The coupling $h_1$ is therefore not running, but
  simply gives a $\Lambda$-independent contribution to the effective potential
  which is the same in both phases.
  It drops
  out when we look at the difference in pressure and we ignore it in the
  remainder of the paper.

      In writing the NLO Lagrangian above, we have ignored
    contributions at finite isospin through the Wess-Zumino-Witten (WZW)
    Lagrangian, which is of the form 
\bqa
{\cal L}_{\text{\tiny WZW}}
\sim\epsilon^{0\mu\nu\alpha}\mu_{I}\Tr\left [\tau_{3}(\Sigma\partial_{\mu}
  \Sigma^{\dagger})(\Sigma\partial_{\nu}\Sigma^{\dagger})
  (\Sigma\partial_{\alpha}
  \Sigma^{\dagger})\right ]\;,
\label{wzw}
\eqa
with the leading contribution at $\mathcal{O}(p^{4})$.
There is also a separate contribution at zero external field at the same
order~\cite{scherer}
but neither of these terms through the WZW action contributes to the
thermodynamic quantities that we compute at one loop.

Expanding the Lagrangian (\ref{lag})
in the fields, we obtain up to quadratic order
  \bqa\nonumber
  \mathcal{L}_{4}^{\rm static}&=&(l_{1}+l_{2})\mu_{I}^{4}\sin^4\alpha
  +l_{4}m^{2}\mu_{I}^{2}\cos\alpha\sin^{2}\alpha
\\ && 
+(l_{3}+l_{4})m^{4}\cos^{2}
  \alpha\;,
  \label{lagstat}
  \eqa
  \bqa\nonumber
  \mathcal{L}_{4}^{\rm linear}&=&(l_{1}+l_{2}) \frac{4\mu_{1}^{4}}{f}
  \cos\alpha
  \sin^{3}
  \alpha\phi_{1}
\\ &&\nonumber
  +l_{4}\frac{m^{2}\mu_{I}^{2}}{f}(2\sin\alpha
  -3\sin^{3}\alpha)\phi_{1}
\\ && \nonumber
  -(l_{3}+l_{4})\frac{2m^{4}}{f}\sin\alpha\cos\alpha\phi_{1}\\
  &&\nonumber
  +(l_{1}+l_{2})\frac{4\mu_{I}^{3}\sin^{3}\alpha}
  {f}
  \partial_{0}\phi_{2}
  \\ 
&&  +l_{4}\frac{2m^{2}\mu_{I}\cos\alpha\sin\alpha}{f}
\partial_{0}\phi_{2}\:,
\label{laglinear}
\eqa
  \bqa
  \nonumber
  \mathcal{L}_{4}^{\rm quadratic}
  &=&(l_{1}+l_{2})\frac{2\mu_{I}^{4}\sin^{2}\alpha}
  {f^{2}} 
  \left [(1+2\cos2\alpha)\phi_{1}^{2}
        \right.\\ &&\left. \nonumber
+\cos^{2}\alpha\phi_{2}^{2}
-\sin^{2}\alpha
    \phi_{3}^{2} \right ]\\ \nonumber
  &&+l_4
  \frac{m^{2}\mu_{I}^{2}\cos\alpha}{4f^{2}}
  \left [(-5+9\cos2\alpha)\phi_{1}^{2}
\right.\\&&\left. \nonumber
    +(1+3\cos2\alpha)\phi_{2}^{2}
    -6\sin^{2}
    \alpha\phi_{3}^{2} \right ]\\ \nonumber
  &&-(l_{3}+l_{4})\frac{m^{4}}{f^{2}}\left [(\cos2\alpha)
    \phi_{1}^{2}
    +\cos^{2}\alpha(\phi_{2}^{2}+\phi_{3}^{2}) \right ]
\\  && \nonumber
  -(l_{1}+l_{2})\frac{4\mu_{I}^{3}\sin\alpha\sin2
    \alpha}
  {f^{2}}(\phi_{2}\partial_{0}\phi_{1}-\phi_{1}\partial_{0}\phi_{2})
\\ \nonumber&& 
-l_{4}\frac{m^{2}\mu_{I}}{f^{2}}
(\cos^{2}\alpha+\cos2\alpha)
    (\phi_{2}\partial_{0}\phi_{1}-\phi_{1}\partial_{0}\phi_{2})
\\ &&\nonumber
+l_{1}\frac{2\mu_{I}^{2}}{f^{2}}\sin^{2}
\alpha(\partial_{\mu}\phi_{a})(\partial^{\mu}\phi_{a})
  \\ &&\nonumber
+l_{2}\frac{2\mu_{I}^{2}}{f^{2}}\sin^{2}
\alpha(\partial_{\mu}\phi_{2})(\partial^{\mu}\phi_{2})
\\&& \nonumber
+(l_{1}+l_2)\frac{4\mu_{I}^{2}\sin^{2}\alpha}{f^{2}}(\partial_{0}\phi_{2})^{2}
\\ && 
  +l_{4}\frac{m^{2}\cos\alpha}{f^{2}}
  (\partial_{\mu}\phi_a)(\partial^{\mu}\phi_a)
  \;,
  \eqa
Eqs.~(\ref{lo1})--(\ref{lo2}) and ~(\ref{cubic})--(\ref{lo4})
from ${\cal L}_2$ and Eq.~(\ref{lagstat}) from ${\cal L}_4$ provide
us with all the terms we need for the NLO calculation within $\chi$PT.
\section{Next-to-leading order  effective potential}
The order-$p^2$ contribution to the effective potential is given by
minus the static part of the Lagrangian ${\cal L}_2$. The one-loop
contribution which is of order $p^4$ is given by a Gaussian path
integral and is ultraviolet divergent. The ultraviolet divergences
must be regularized and we choose dimensional regularization.
Dimensional regularization sets
power divergences to zero and logarithmic divergences show
up as poles in $\epsilon$, where $d=3-2\epsilon$ is the number of spatial
dimensions (see below). The divergences are cancelled by renormalizing
the coupling constants appearing in the static part of the Lagrangian
${\cal L}_4$, which is also of order-$p^4$.
\subsection{Vacuum phase}
The order-$p^2$ contribution $V_0$ to the effective potential $V_{\rm eff}$
is equal to minus the static Lagrangian given in Eq.~(\ref{lo1}),
evaluated at $\alpha=0$,
\bqa
V_0&=&-f^2m^2\;.
\eqa
The dispersion relations for the neutral pion reduces to
$E_{\pi^0}=\sqrt{p^2+m^2}$ and for the charged pions
$E_{\pi^{\pm}}=\sqrt{p^2+m^2}\mp\mu_I$. The one-loop contribution to the effective
potential is therefore
\bqa\nonumber
V_1&=&V_{1,\pi^0}+V_{1,\pi^+}+V_{1,\pi^-}
={1\over2}\int_p\left(E_{\pi^0}+E_{\pi^{+}}+E_{\pi^{-}}\right)
\\
&=&
{3\over2}\int_p\sqrt{p^2+m^2}\;.
\eqa
The integral is defined as 
\bqa
\int_{p}&=&\left (\frac{e^{\gamma_{E}}\Lambda^{2}}{4\pi} \right )^{\epsilon}
\int
\frac{d^{d}p}{(2\pi)^{d}}\;,
\label{sumint}
\eqa
where $\Lambda$ is the renormalization scale in the modified minimal
subtraction ($\overline{\rm MS}$) scheme and $d=3-2\epsilon$
is the number of spatial
dimensions. Using Eq.~(\ref{int1}), we find
\bqa
V_1&=&
-{3m^4\over4(4\pi)^2}
\left [\frac{1}{\epsilon}+\frac{3}{2} +
  \log\left (\frac{\Lambda^{2}}{m^{2}} \right )\right ]\;.
\eqa
The ${\cal O}(p^4)$ static term $V_1^{\rm static}$
is given by minus ${\cal L}_4^{\rm static}$
evaluated at $\alpha=0$,
\bqa
V_1^{\rm static}&=&-(l_3+l_4)m^4\;.
\eqa
Using Eq.~(\ref{lowl}) with $i=3,4$, 
the renormalized one-loop effective potential is then given by
\bqa\nonumber
V_{\rm eff}&=&V_0+V_1^{\rm static}+V_1
\\
&=&
-f^2m^2
-{3m^4\over4(4\pi)^2}
\left [\frac{1}{2} 
-{1\over3}\bar{l}_3+{4\over3}\bar{l}_4
\right ]
\;.
\label{vacu}
\eqa
We note that Eq.~(\ref{vacu}) and therefore
the thermodynamic quantities are independent of the
isospin chemical potential $\mu_I$ all the way up to $\mu_I=m_{\pi}$
(see Sec.~\ref{numres}), which is the Silver-Blaze property~\cite{cohen}.
We therefore refer to this as the vacuum phase.
The scale dependence has cancelled in the final result Eq.~(\ref{vacu}).

\subsection{Pion-condensed phase}
The order-$p^2$ contribution $V_0$ to the effective potential 
$V_{\rm eff}$ is equal to minus the static Lagrangian given in Eq.~(\ref{lo1}),
\bqa
V_0&=&-f^{2}m^{2}\cos\alpha-
  {1\over2}f^2\mu_{I}^{2}\sin^{2}\alpha\;.
\eqa
Using the dispersion relations for the pions, we can write down the one-loop
contribution to the effective potential as follows
\bqa
\nonumber
V_{\rm 1}&=&V_{1,\pi^0}+V_{{\rm 1}, \pi^{+}}
+V_{{\rm 1}, \pi^{-}}
=
\frac{1}{2}\int_{p}E_{\pi^0}+
\frac{1}{2}\int_{p}(E_{\pi^+}+E_{\pi^-})\;,
\\ &&
\label{f11}
\eqa
Using Eq.~(\ref{int1}), we find
\bqa\nonumber
\nonumber
V_{{\rm 1},\pi^0}&=&\frac{1}{2}\int_{p}\sqrt{p^{2}+m_{3}^{2}}
=%
-\frac{m_{3}^{4}}{4(4\pi)^{2}}\left [\frac{1}{\epsilon}+\frac{3}{2} +
  \log\left (\frac{\Lambda^{2}}{m_{3}^{2}} \right )\right ]\;.
\\&&
\label{lastcon}
\eqa
The calculation of $V_{1,\pi^{\pm}}$
requires isolating the ultraviolet divergences,
which can be done by expanding $E_{\pi^{\pm}}$
in powers of $\frac{1}{p}$, which
gives
\bqa\nonumber
E_{\pi^+}+E_{\pi^-}&=&
2p+\frac{2(m_{1}^{2}+m_{2}^{2})+m_{12}^{2}}{4p}
\\ && \nonumber
-\frac{8(m_{1}^{4}+m_{2}^{4})+4(m_{1}^{2}+m_{2}^{2})m_{12}^{2}
  +m_{12}^{4}}{64p^{3}}+...
\\ &&
\eqa
The ultraviolet behavior of $E_{\pi^+}+E_{\pi^-}$ 
is the same as that of $E_{1}+E_{2}$, where
$E_{i}=\sqrt{p^2+m_i^2+\mbox{$1\over4$}m^4_{12}}$ ($i=1,2$).
Defining
$\tilde{m}_1^2=m_1^2+\mbox{$1\over4$}m^4_{12}=
m^2\cos\alpha+\mu_I^2\sin^2\alpha=m_3^2$
and
$\tilde{m}_2^2=m_2^2+\mbox{$1\over4$}m^4_{12}=m^2\cos\alpha$,
the divergent part of the first two terms in Eq.~(\ref{f11})
reads
\bqa\nonumber
V^{\rm div}_{{\rm 1},\pi^+}+V^{\rm div}_{{\rm 1},\pi^-}
&=&-
\frac{\tilde{m}_1^4}{4(4\pi)^{2}}
\left [\frac{1}{\epsilon}+{3\over2}
  + \log\left (\frac{\Lambda^{2}}{\tilde{m_1}^2}\right )\right ]
\\&&-
\frac{\tilde{m}_2^4}{4(4\pi)^{2}}
\left [\frac{1}{\epsilon}+{3\over2}
+ \log\left (\frac{\Lambda^{2}}{\tilde{m_2^2}}\right )\right ]\;,
\label{divv1}
\eqa
where we have used Eq.~(\ref{int1}).
The finite part is defined as
\bqa
V_{{\rm 1},\pi^+}^{\rm fin}+V_{{\rm 1},\pi^-}^{\rm fin}
&=&\frac{1}{2}\int_{p}\left [E_{\pi^+}+E_{\pi^-}
-E_1-E_2
\right ]\;,
\label{finv1}
\eqa
such that the sum of Eqs.~(\ref{divv1}) and Eqs.~(\ref{finv1})
is equal to the first two terms in Eq.~(\ref{f11}).
The expression for the divergent pieces can be written in terms of $\alpha$
using the explicit expressions for $m_{i}$, Eqs.~(\ref{m1def})--(\ref{m4def}).
We find
\bqa\nonumber
V_{{\rm 1}}^{\rm div}
&=&
-\frac{1}{2(4\pi)^{2}}
\left [\frac{1}{\epsilon}+\frac{3}{2} +
  \log\left (\frac{\Lambda^{2}}{{m}_{3}^{2}} \right )\right ]
\\ && \nonumber
\times
\left(m^2\cos\alpha+\mu_{I}^{2}\sin^{2}\alpha\right)^2\\
&&-\frac{1}{4(4\pi)^{2}}
\left [\frac{1}{\epsilon}+\frac{3}{2} +
  \log\left (\frac{\Lambda^{2}}{\tilde{m}_{2}^{2}} \right )\right ]
m^4\cos^2\alpha\;.
\eqa
The static ${\cal O}(p^4)$ comes from the static part of the Lagrangian,
given by minus Eq.~(\ref{lagstat}),
\bqa\nonumber
V_1^{\rm static}
&=&-(l_{1}+l_{2})\mu_{I}^{4}\sin^4\alpha
  -l_{4}m^{2}\mu_{I}^{2}\cos\alpha\sin^{2}\alpha
\\ &&-(l_3+l_4)m^4\cos^2\alpha\;.
  \eqa
After renormalization, using Eq.~(\ref{lowl})
the effective potential $V_{\rm eff}=V_0+V_1^{\rm static}+V_1$
has the form
\bqa\nonumber
V_{\rm eff}&=&-f^{2}m^{2}\cos\alpha-\frac{1}{2}f^{2}\mu_{I}^{2}
\sin^{2}\alpha
\\ && \nonumber
-\frac{3}{4(4\pi)^{2}}\left [\frac{1}{2}-{1\over3}\bar{l}_{3}
  +{4\over3}\bar{l}_{4}
  +{1\over3}\log\left({m^2\over \tilde{m}_2^2}\right)
\right.\\ &&\left. \nonumber
  +  {2\over3}\log\left({m^2\over m_3^2}\right)
\right ]
m^{4}\cos^{2}\alpha
\\&& \nonumber
-\frac{1}{(4\pi)^{2}}
\left [{1\over2}+\bar{l}_{4}
  +  \log\left({m^2\over m_3^2}\right)
\right ]
m^{2}\mu_{I}^{2}\cos\alpha\sin^{2}\alpha
\\ &&\nonumber
-\frac{1}{2(4\pi)^{2}}\left [{1\over2}
  +\frac{1}{3}\bar{l}_{1}+\frac{2}{3}\bar{l}_{2}
  +  \log\left({m^2\over m_3^2}\right)
\right ]\mu_{I}^{4}\sin^{4}\alpha
\\ &&
+V_{{\rm 1},\pi^+}^{\rm fin}+V_{{\rm 1},\pi^-}^{\rm fin}
\;.
\label{effpotnlo}
\eqa
We note that the all the $\Lambda$-dependence cancels in the final
result (\ref{effpotnlo}). This implies that the thermodynamic functions
are independent of the renormalization scale.

\section{Thermodynamics}
\label{numres}
In this section, we investigate the thermodynamics
of the pion-condensed phase using the
effective potential~(\ref{effpotnlo}).
We will calculate the pressure ${\cal P}$ and
the isospin density $n_I$ as a function of the isospin
chemical potential $\mu_{I}$, as well
as the equation of state, i.e. the energy density $\epsilon$ 
as a function of the pressure ${\cal P}$.
In order to evaluate these quantities we need to know the low-energy
constants $\bar{l}_i$. Evaluated at the scale $\mu=m$, they have the following
values and uncertainties~\cite{cola}
\begin{align}
  \label{variasjon1}
\bar{l}_{1}&=-0.4\pm 0.6\;,
&\bar{l}_{2}=4.3\pm 0.1\;,\\
\bar{l}_{3}&=2.9 \pm 2.4\;,
&\bar{l}_{4}=4.4 \pm 0.2\;.
\label{variasjon2}
\end{align}
The coupling constants $\bar{l}_1$ and $\bar{l}_2$ can be measured
experimentally via the $d$-wave scattering lengths, while
the coupling constant $\bar{l}_3$ has been estimated using three-flavor
QCD \cite{gasser1}. Finally, the coupling $\bar{l}_4$ is related to
the scalar radius of the pion and has also been estimated to the value
quoted above.

At LO, $m=m_{\pi}$ and 
$f=f_{\pi}$ and so their uncertainties are the same.
Given the values of $\bar{l}_3$ and $\bar{l}_4$, the parameters
$m^2$ and $f^2$ at NLO are determined using Eqs.~(\ref{mpi}) and ~(\ref{fpi})
and the values for the pion mass
and the pion decay constant. 
Since we want to compare our results to lattice data, we choose
the same pion mass and pion decay constant~\cite{gergy11},
\begin{align}
m_{\pi}&=131\pm3 {\rm MeV}\;,
&f_{\pi}={128\pm3\over\sqrt{2}}{\rm MeV}\;.
\end{align}
The uncertainties in the low-energy constants, $m_{\pi}$,
and $f_{\pi}$ translate into
uncertainties in $m$ and $f$.
The central values $m_{\rm cen}$
and $f_{\rm cen}$ are obtained by using the central values of
$\bar{l}_i$, $m_{\pi}$ and $f_{\pi}$. 
The
minimum and maximum values of
$m$ and $f$ denoted by
$m_{\min},\ f_{\min}$ and $m_{\max},\ f_{\max}$ respectively are
obtained by combining
the maximum and minimum values of the $\bar{l}_i$s, $f_{\pi}$, and $m_{\pi}$.
The values for the bare pion mass and decay constant are
\begin{align}
\label{pp1}
  m_{\rm cen}&=132.4884\,{\rm MeV}\;,
&f_{\rm cen}=84.9342\,{\rm MeV}\;,\\
m_{\rm min}&=128.2409\,{\rm MeV}\;,
&f_{\rm min}=83.2928\,{\rm MeV}\;, \\
m_{\rm max}&=136.9060\,{\rm MeV}\;,
&f_{\rm max}=86.5362\,{\rm MeV}\;.
\label{pp3}
\end{align}
We have also considered separately the uncertainties in the LECs and
the parameters $m_{\pi}$ and $f_{\pi}$.
It turns out that the uncertainties are completely
dominated by the latter.

The thermodynamic functions
are derived from the effective potential~(\ref{effpotnlo})
at its minimum as a function of $\alpha$ so we must first solve
\bqa
{\partial V_{\rm eff}\over\partial\alpha}&=&0\;.
\label{alphi1}
\eqa
This can also be used to show that the linear term vanishes on-shell i.e.
for the value of $\alpha$ that minimizes $V_{\rm eff}$.
We show this explicitly in \ref{eosapp}.

\begin{figure}[htb]
  \includegraphics[width=0.5\textwidth]{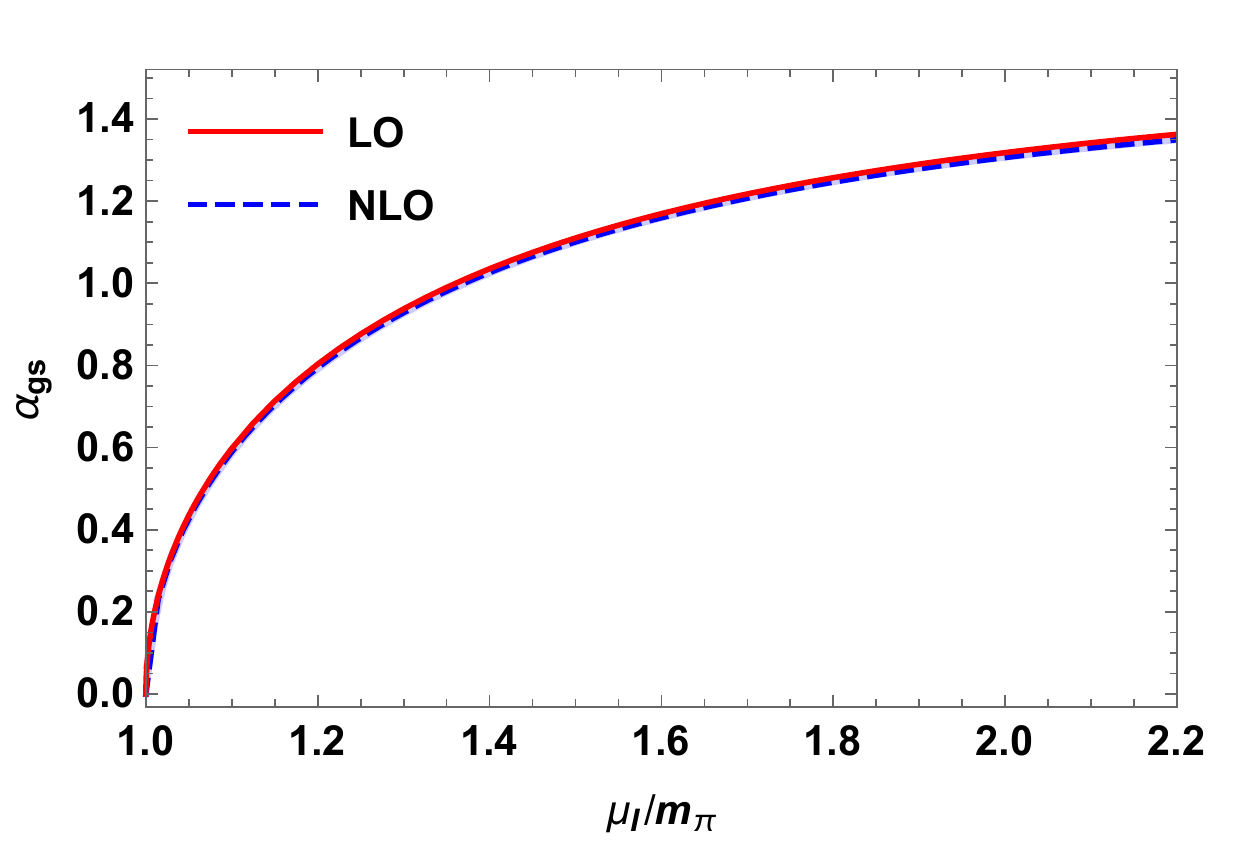}
  \caption{$\alpha$ that minimizes the effective potential as a function of
    isospin chemical potential $\mu_I$. The red curve is the LO
    results, while the blue curve is the NLO result.
  }
\label{alphi}
\end{figure}

In Fig.~\ref{alphi}, we show the solution to Eq.~(\ref{alphi1}) as
function of the isospin chemical potential $\mu_I$
divided by $m_{\pi}$. The red curve is
the order-$p^2$ result, while the blue curve is the order-$p^4$ result.
The curves are barely distinguishable.

\begin{figure}[htb]
  \includegraphics[width=0.5\textwidth]{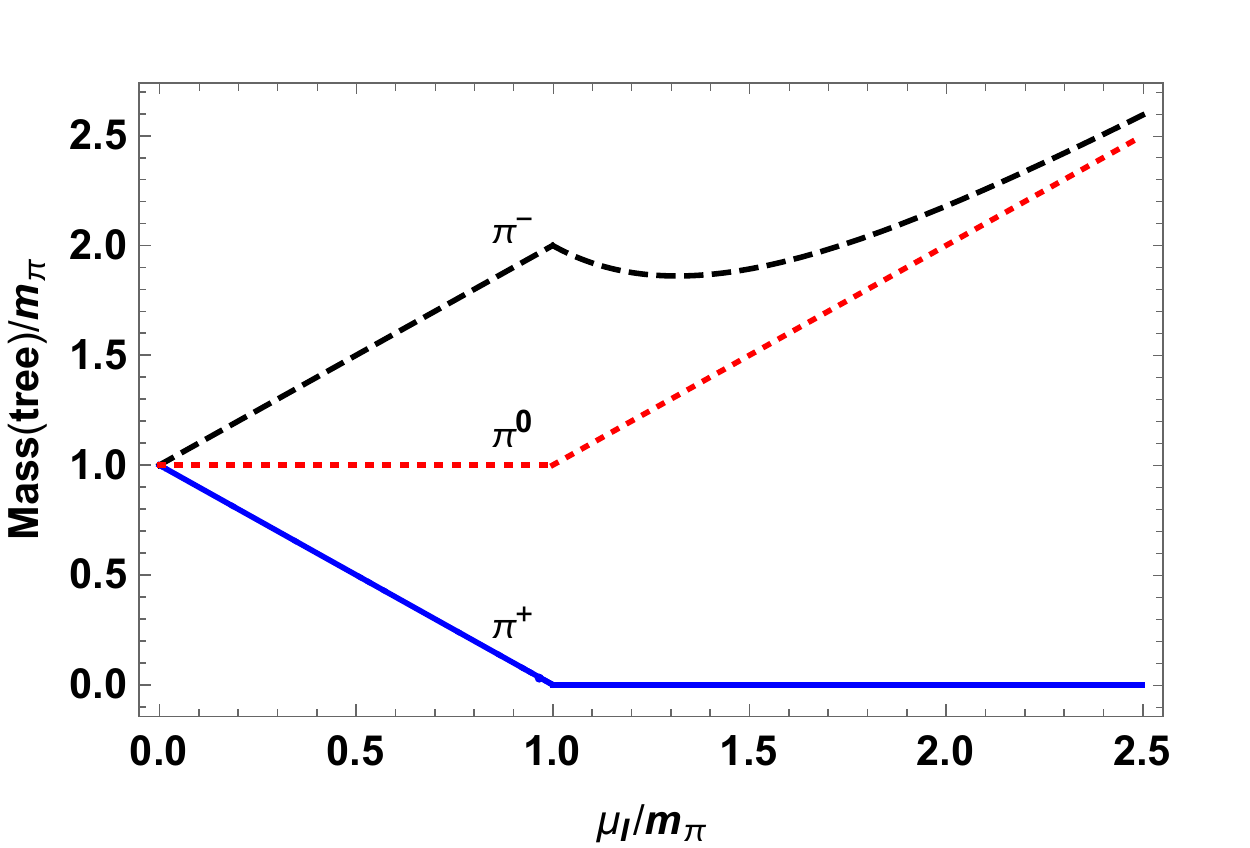}
  \caption{Tree-level masses normalized to the pion mass in the vacuum
    as a function of isospin chemical potential normalized
  by the pion mass in the vacuum.}
\label{masses}
\end{figure}

We first discuss the quasi-particle masses. Restricting ourselves to
tree level, the masses are obtained by setting $p=0$
in Eqs.~(\ref{pipo})--(\ref{pipo2}).
The normalized masses are shown in Fig.~\ref{masses}
as a function of the normalized isospin chemical potential (both normalized
by the pion mass in the vacuum). The mass of the
neutral pion is given by the red dotted line,
the black curve is the mass of $\pi^-$, and the blue line is the
mass of $\pi^+$.

We see that the pionic excitation $\pi^+$ is massless for $\mu_I\geq m_{\pi}$,
In the pion-condensed phase, $m^2_2=0$ at the minimum
of the effective potential.
Expanding Eq.~(\ref{pipo}) around $p=0$ yields
\bqa
E_{\pi^+}&=&\sqrt{\mu_I^4-m_{\pi}^4\over3m_{\pi}^4+\mu_I^4}p+{\cal O}(p^2)\;,
\eqa
where we have set $m=m_{\pi}$ which is correct at LO.
This shows explicitly
that $\pi^+$ is a massless excitation, which arises due to
spontaneous breaking of the $U(1)_{I_3}$ symmetry in the pion-condensed phase.

In order to show that there is a second-order transition at
a critical chemical potential $\mu_I^c=m_{\pi}$, we
expand the effective potential in powers of $\alpha$ up to
$\mathcal{O}(\alpha^{4})$ to obtain an effective Landau-Ginzburg
energy functional, 
\bqa
\label{LG}
V_{\rm eff}^{\rm LG}=a_0+a_2(\mu_I)\alpha^2+a_4(\mu_I)\alpha^4\;.
\eqa
In \ref{alpharekkje}, we carry out the expansion of the effective potential
to order $\alpha^4$ using the techniques Ref.~\cite{split2}.
The coefficient $a_2(\mu_I)$ can be read off from Eq.~(\ref{readoff}), 
\bqa
a_2(\mu_I)&=&{1\over2}f_{\pi}^2\left[m_{\pi}^2-\mu_I^2\right]\;.
\eqa
The critical isospin chemical potential $\mu_I^c$ is
defined by the vanishing of the coefficient of the
$\alpha^2$ term, i.e. $a_2(\mu_I^c)=0$.
This shows that $\mu_I^c=m_{\pi}$.
In order to obtain this result, we had to take into account the
one-loop corrections to the pole mass of the pion and to the pion
decay constant 
expressed
in terms of $m$, $f$ and the low-energy constants, cf.
Eqs.~(\ref{mpi})--(\ref{fpi}).
This result holds to all orders in perturbation theory and is
also in agreement with the lattice simulations of~\cite{gergy1,gergy2,gergy3}.
Moreover,
if $a_4(\mu_I^c)>0$, the transition is second order.
The coefficient $a_4(\mu_I)$ can be read off from Eq.~(\ref{readoff}).
Evaluated at $\mu_I=m_{\pi}$, we find
\bqa\nonumber
a_4(\mu_I^c)
&=&\frac{1}{8}f^{2}m^{2}\left\{
1-{m^2\over2(4\pi)^2f^2}\left[
1+{8\over3}\bar{l}_{1}+
{16\over3} \bar{l}_{2}-8\bar{l}_{4}\right]\right\}\;,\\
\label{LG2}
\eqa
which is larger than zero. This
means that the onset of pion condensation 
is via a second-order transition exactly at the physical pion mass.

\begin{figure}[htb]
  \includegraphics[width=0.5\textwidth]{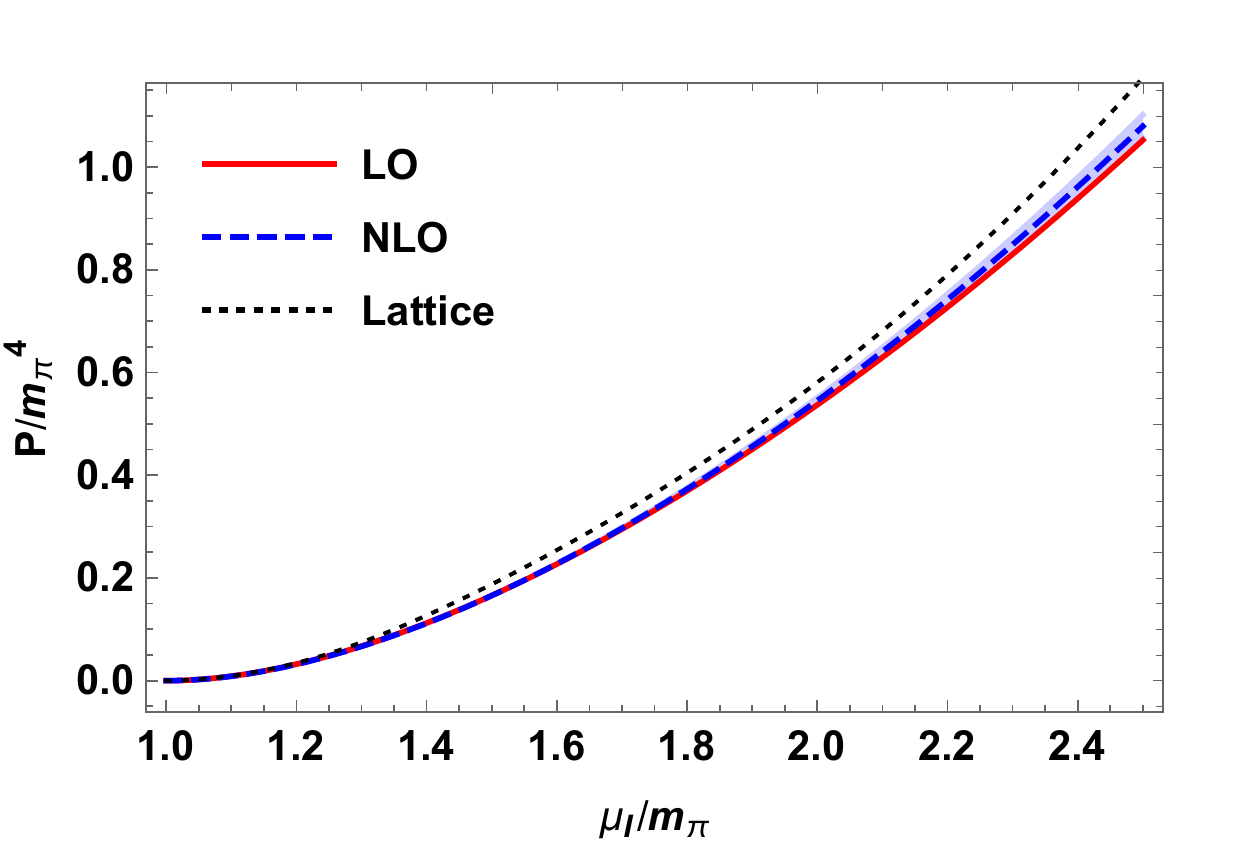}
  \caption{The normalized pressure as a function of the normalized
    isospin chemical potential.
    The tree-level and one-loop results are the red solid and blue dashed line,
    respectively, using $m_{\rm cen}$ and $f_{\rm cen}$. The band is obtained
    by varying $m$ and $f$ in their respective ranges.
    The dashed line is the lattice results from Ref.~\cite{endro}.}
\label{P}
\end{figure}

We next turn to the thermodynamic functions.
The pressure is given by ${\cal P}=-V_{\rm eff}$. Since we are interested
in the pressure relative to the vacuum phase we subtract the
pressure for $\alpha=0$, and define
\bqa
{\cal P}&=&-V_{\rm eff}+V_{\rm eff}(\alpha=0)\;,
\label{pdiff}
\eqa
  where the effective potential is evaluated at the minimum.
In Fig.~\ref{P}, we show the
pressure normalized to $m_{\pi}^4$
as a function of the isospin chemical potential normalized to $m_{\pi}$.
The red curve is the leading-order result, while the blue curve is
the next-to-leading order result using the central values
of $m$ and $f$.
The NLO band is obtained by varying the 
parameters of $m$ and $f$ as given in Eqs.~(\ref{pp1})--(\ref{pp3}).
We also show the lattice results for the pressure from Ref.~\cite{endro}.
The pressure increases steadily with the
    chemical potential. The NLO pressure increases faster than
    the LO pressure and is in good agreement with the lattice results.
    
\begin{figure}[htb]
  \includegraphics[width=0.5\textwidth]{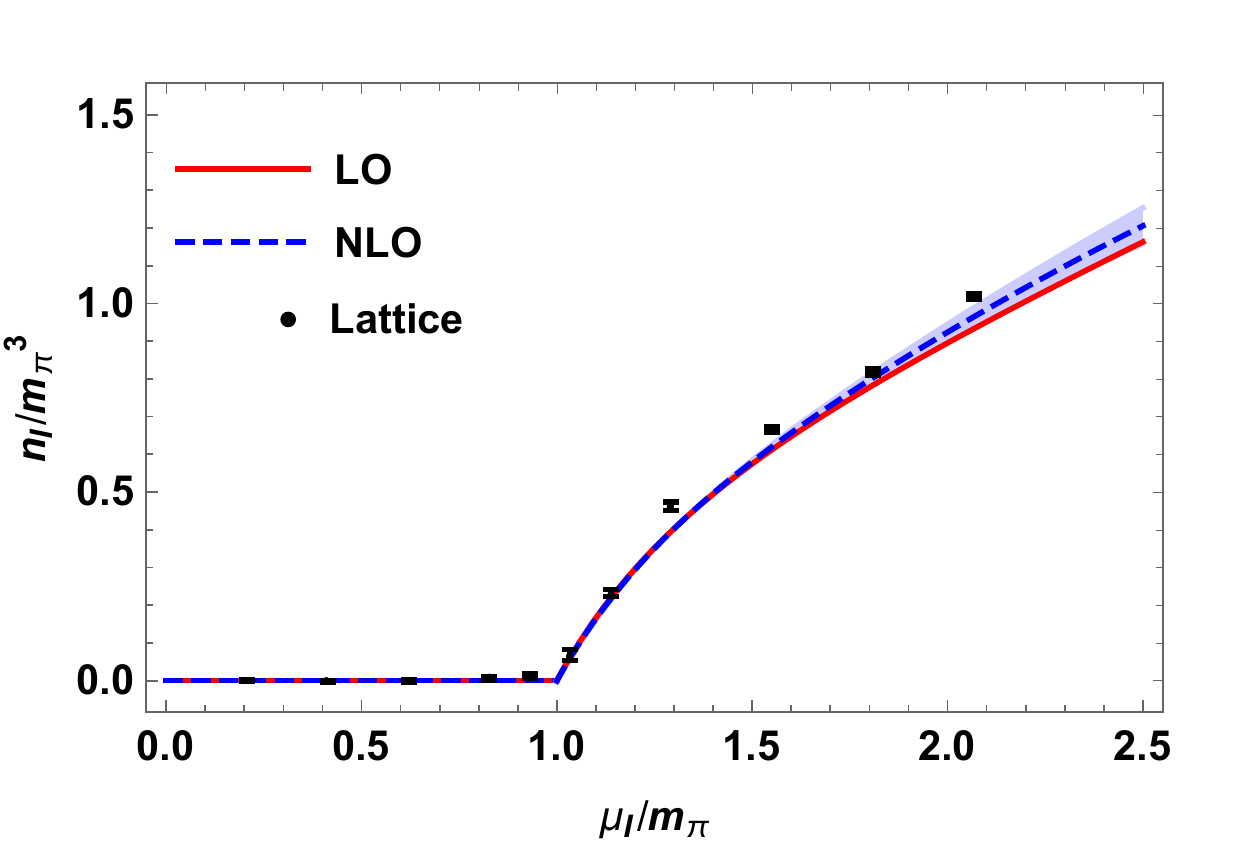}
  \caption{Normalized isospin density as a function of the
    normalized isospin chemical potential. 
    The red curve shows the
    tree-level result and the blue curve shows the one-loop result using the
    central values $m_{\rm cen}$ and $f_{\rm cen}$.
    The band is obtained by varying $m$ and $f$ in their respective ranges.
    The points are lattice data from Ref.~\cite{endro}.
  }
\label{nI}
\end{figure}

The isospin density is defined as
\bqa\nonumber
n_{I}&\equiv&-\frac{\partial V_{\rm eff}}{\partial\mu_{I}}\\
&=& \nonumber
f^2\mu_I\sin^2\alpha
+{2\over(4\pi)^2}\left[
  \bar{l}_4+\log{m^2\over m_3^2}\right]
m^2\mu_I\cos\alpha\sin^2\alpha
\\ \nonumber
&&+{2\over(4\pi)^2}\left[
  {1\over3}\bar{l}_1+{2\over3}\bar{l}_2+
  \log{m^2\over m_3^2}\right]\mu_I^3\sin^4\alpha
\\&& 
-{\partial(V_{{\rm 1},\pi^+}^{\rm fin}+V_{{\rm 1},\pi^-}^{\rm fin})
  \over\partial\mu_I}
\;.
\eqa
In Fig.~\ref{nI}, we show the isospin density normalized by $m_{\pi}^3$
as a function of the chemical potential $\mu_I$ normalized by
$m_{\pi}$.
The red curves shows the
tree-level result and the blue curve shows the one-loop result
using the central values of the parameters $m$ and $f$.
The band is obtained by varying the 
parameters $m$ and $f$ as given by Eqs.~(\ref{pp1})--(\ref{pp3}).
We also show the lattice points from Ref.~\cite{endro}.
There is no pion condensate in the vacuum up to the critical
isospin chemical potential $\mu_I^c=m_{\pi}$. Hence $n_I$ is independent of
$\mu_I$, which is an example of the Silver-Blaze property, namely
that thermodynamic functions do not depend on $\mu_I$ all the way up to
its critical value~\cite{cohen}. For $\mu_{I}$ larger than the critical isospin
chemical potential $\mu_I^c=m_{\pi}$, the density increases steadily.
The isospin density as a function of $\mu_I$ increases as one goes
from LO to NLO, and the latter is in better agreement with
the lattice results of Ref.~\cite{endro}.

\begin{figure}[htb]
  \includegraphics[width=0.5\textwidth]{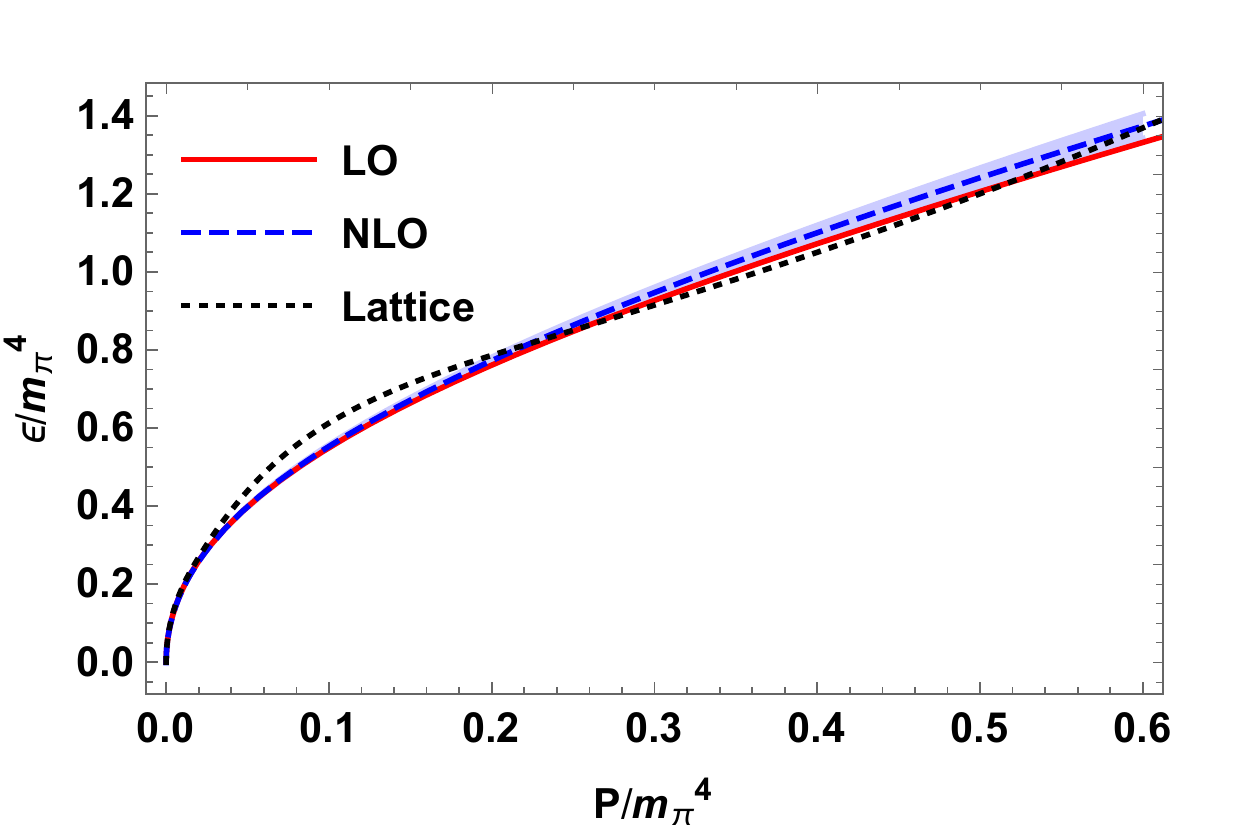}
  \caption{The normalized equation of state 
    at tree level is the red curve and at one loop is the 
    blue curve using  the central values $m_{\rm cen}$ and $f_{\rm cen}$.
    The blue band is obtained by varying the parameters $m$ and $f$
    in their respective ranges.
    The dashed line is the lattice results from Ref.~\cite{endro}.}
\label{eos}
\end{figure}

The energy density is defined  by
\bqa\nonumber
\epsilon&=&-{\cal P}+n_I\mu_I
\\ &=& \nonumber
-V_{\rm eff}(\alpha=0)-f^2m^2\cos\alpha
+{1\over2}f^2\mu_I^2\sin^2\alpha
\\ &&\nonumber
-\frac{3}{4(4\pi)^{2}}\left [\frac{1}{2}-{1\over3}\bar{l}_{3}
  +{4\over3}\bar{l}_{4}
  +{1\over3}\log\left({m^2\over \tilde{m}_2^2}\right)
\right.\\ && \left.\nonumber
+  {2\over3}\log\left({m^2\over m_3^2}\right)
\right ]m^{4}\cos^{2}\alpha
\\ && \nonumber
-{1\over(4\pi)^2}\left[{1\over2}-\bar{l}_4
  -\log{m^2\over m_3^2}\right]
m^2\mu_I^2\cos\alpha\sin^2\alpha
\\ \nonumber
&&-{1\over2(4\pi)^2}\left[{1\over2}
  -\bar{l}_1-2\bar{l}_2-
  3\log{m^2\over m_3^2}\right]
\mu_I^4\sin^4\alpha
\\
&&
+V_{{\rm 1},\pi^+}^{\rm fin}+V_{{\rm 1},\pi^-}^{\rm fin}
-\mu_I{\partial(V_{{\rm 1},\pi^+}^{\rm fin}+V_{{\rm 1},\pi^-}^{\rm fin})
  \over\partial\mu_I}
\;,
\eqa
and can be used to find the EoS. 
In Fig.~\ref{eos}, we show the normalized equation of state.
The LO result is the red curve while the NLO result is the blue curve
using the central values of the parameters $m$ and $f$. 
The blue band is obtained by varying the 
parameters of $m$ and $f$ as given by Eqs.~(\ref{pp1})--(\ref{pp3}).
The black dashed line shows the lattice results from Ref.~\cite{endro}.
We notice that the NLO equation of state is stiffer than the LO one and that
the difference increases steadily with the pressure ${\cal P}$. Moreover,
the NLO EoS is in better agreement with the lattice results
for small values of ${\cal P}/m_{\pi}^4$ than the LO
EoS, while for larger values it is the other way around.

\section{Summary}
In conclusion, we have derived the $\chi$PT Lagrangian 
which is necessary for all NLO calculations at finite isospin.
We have applied this Lagrangian calculating the pressure, isospin
density, as well as the equation of state.
Our predictions are in good agreement with the lattice results of
Ref.~\cite{endro} and improves as one goes from LO to NLO.
This is the first test of $\chi$PT
in the pion-condensed phase 
beyond leading order.
The Lagrangian we have derived 
can be used to calculate e.g. the one-loop corrections
to the quasiparticle masses in the pion-condensed phase. Here a nontrivial
check would be to show that one of the branches is a massless Goldstone
boson. The Lagrangian
for three-flavor QCD can be derived in the same way and opens up the possibility
to study quantum effects in phases that involve pion or kaon condensation.
In the case of pion condensation, one can again compare with the
lattice results of Ref.~\cite{endro}, as well as between those of the two
and three-flavor calculations. This will give us an idea of the
effects of the strange quark.
Work in this direction is in progress~\cite{us}.

\section*{Acknowledgements}
The authors would like to thank B. Brandt, G. Endr\H{o}di and S. Schmalzbauer
for useful discussions as well as for providing the
data points of Ref. \cite{endro}. The authors would also like to
thank the Niels Bohr International Academy for hospitality during the
later stages of this work.
P. A. would like to acknowledge the Faculty Life Committee at St. Olaf College
and the Nygaard Study in Norway Endowment for partial travel support.

\appendix

\section{Dimensionally regularized integrals}
We need a single integral in $d=3-2\epsilon$ dimensions,
\bqa
\label{int1}
\int_p\sqrt{p^2+m^2}&=&
-{m^4\over2(4\pi)^2}
\left({{\Lambda^2\over m^2}}\right)^{\epsilon}
\left[{1\over\epsilon}+{3\over2}+{\cal O}(\epsilon)\right]\;.
\eqa
We need several integrals in $d+1=4-2\epsilon$ dimensions.
The integrals are defined as in Eq.~(\ref{sumint}), execpt
that the integral is over $P$ in $d+1$ Euclidean dimensions.
\begin{align}
\label{standard1}  
\int_P\log\left[P^2+m^2\right]
=-{m^4\over2(4\pi)^2}
\left({{\Lambda^2\over m^2}}\right)^{\epsilon}
\left[{1\over\epsilon}+{3\over2}+{\cal O}(\epsilon)
\right]&\;,\\
\label{d4}
\int_P{1\over P^2+m^2}=-{m^2\over(4\pi)^2}
\left({{\Lambda^2\over m^2}}\right)^{\epsilon}
\left[{1\over\epsilon}+1+{\cal O}(\epsilon)
\right]&\;,\\ \nonumber
\int_P{1\over[p^2+{1\over2}(m_1^2+m_2^2)^2]^2+p_0^2m_{12}^2}
  ={1\over(4\pi)^2}\bigg[{1\over\epsilon}+2
    \\ \nonumber
  -
2\log{\sqrt{m_1^2+m_2^2}+\sqrt{m_1^2+m_2^2+m_{12}^{2}}\over2\Lambda}
  +{2(m_1^2+m_2^2)\over m_{12}^2}&
  \\
\label{standard2}    
  -{2\sqrt{(m_1^2+m_2^2)(m_1^2+m_2^2+m_{12}^{2})}\over m_{12}^2}
+{\cal O}(\epsilon)\bigg]&\;.
\end{align}

The integrals (\ref{standard1}) and (\ref{d4}) are standard, while
details of the evaluation of (\ref{standard2}) can be found in
Ref.~\cite{split2}.

\section{Mass renormalization}
\label{relations}
In order to show the second-order nature of
the phase transition to a Bose-condensed phase at $\mu_I^c=m_{\pi}$,
where $m_{\pi}$ is the physical mass in the vacuum, we
need to express it in terms
of the parameters $m$ and $f$ of the chiral Lagrangian.
The relevant terms are found by setting $\alpha=0$
in the ${\cal L}_2$ and ${\cal L}_4$, both evaluated at $\mu_I=0$,
\bqa\nonumber
\mathcal{L}_{2}^{\rm quartic}&=&\frac{m^2}{24f^{2}}(\phi_{a}
\phi_{a})^2
  +\frac{1}{6f^{2}}\left [\phi_{a}\phi_{b}
    \partial^{\mu}
    \phi_{a}    \partial^{\mu}\phi_{b}
\right.\\ &&\left.
    -\phi_{a}\phi_{a}\partial_{\mu}
    \phi_{b}
    \partial^{\mu}\phi_{b} \right ]\;,\\
  \mathcal{L}_{4}^{\rm quadratic}
  &=&
  -(l_{3}+l_{4})\frac{m^{4}}{f^{2}}\phi_a\phi_a
  +l_{4}\frac{m^{2}}{f^{2}}
    (\partial_{\mu}\phi_a)(\partial^{\mu}\phi_{a})
    \;.  
\eqa
The inverse propagator for the pion is
\bqa
\label{inverse}
K^2-m^2-\Sigma_1(K^2)-\Sigma_{\rm 2}(K^2)\;,
\eqa
where the ${\cal O}(p^2)$ self-energy in the vacuum is 
\bqa
\Sigma_1(K^2)&=&-{2iK^2\over3f^2}\int_P{1\over P^2-m^2}
+{im^2\over6f^2}\int_P{1\over P^2-m^2}\;,\\
\Sigma_{\rm 2}(K^2)&=&2K^2{m^2\over f^2}l_4
-{2m^4\over f^2}(l_3+l_4)
\;.
\eqa
Here the integral is in Minkowski space. 
The physical pion mass $m_{\pi}$ is defined as the pole of the
propagator, or
\bqa
m_{\pi}^2-m^2-\Sigma_1(m_{\pi}^2)-\Sigma_{\rm 2}(m_{\pi}^{2})
&=&0\;.
\eqa
Solving this equation self-consistently to NLO and going to Euclidean
space yield
\bqa\nonumber
m_{\pi}^2&=&m^2+{m^2\over2f^2}\int_P{1\over P^2+m^2}
+{2m^4\over f^2}l_3\\
&=&m^2\left[1-{m^2\over2(4\pi)^2f^2}\bar{l}_3\right]\;,
\label{mpi}
\eqa
where we have used Eq.~(\ref{lowl}) with $i=3$, 
and Eq.~(\ref{d4}).
The pion decay constant $f_{\pi}$
can be determined in a similar manner, either through the coupling of the
axial current to the pion, or by calculating the correlator between two
axial currents. The result is~\cite{gasser1}
\bqa
f_{\pi}^2
&=&f^2\left[1+{2m^2\over(4\pi)^2f^2}\bar{l}_4\right]\;.
\label{fpi}
\eqa

\section{Rotated generators}
\label{rotgen}
Let us consider the rotated parametrization
$L_{\alpha}$ given by 
\bqa
L_{\alpha}&=&A_{\alpha}UA_{\alpha}^{\dagger}\;.
\eqa
An infinitesimal fluctuation can be written as
\bqa\nonumber
L_{\alpha}&=&
\left[
  \cos{\alpha\over2}+i\tau_1\sin{\alpha\over2}\right]
\left[
1+i{\phi_a\tau_a\over2f}
\right]
\left[
  \cos{\alpha\over2}-i\tau_1\sin{\alpha\over2}\right] \\ \nonumber
&=&
1+{i\phi_1\tau_1\over2f}+{i\phi_2\over2f}
(\cos\alpha\tau_2-\sin\alpha\tau_3)
\\
&&+{i\phi_3\over2f}(\cos\alpha\tau_3+\sin\alpha\tau_2)\;.
\label{trafo}
\eqa
We can define new rotated generators 
$\tau_i^{\prime}$ as
\bqa
\tau_1^{\prime}&=&\tau_1\;,\\
\tau_2^{\prime}&=&(\cos\alpha\tau_2-\sin\alpha\tau_3)\;,\\
\tau_3^{\prime}&=&(\cos\alpha\tau_3+\sin\alpha\tau_2)\;.
\eqa
It is easy to show that the generators $\tau_i^{\prime}$
satisfy the standard commutation relations of the Pauli matrices.
The form of the rotated generators can be understood as follows.
The vacuum is rotated in the plane spanned by $\mathbb{1}$ and
$\tau_1$, which implies that only the generators in the other
directions, i.e. $\tau_2$ and $\tau_3$ are rotated.
To all orders in $\alpha$, we can write
\bqa
L_{\alpha}&=&\exp\left(i{\phi_a\tau^{\prime}_a\over2f}\right)\;.
\eqa
\section{Equation of motion}
\label{eosapp}
The equation of motion for the effective potential in the absence of
sources is
\bqa
{\partial V_{\rm eff}\over\partial\alpha}=0\;.
\eqa
At tree level, the potential $V_0$ is given by minus
Eq.~(\ref{lo1}). Minimizing $V_0$ yields
$fm^2\sin\alpha-f\mu_I^2\sin\alpha\cos\alpha=0$.
Comparing Eq.~(\ref{lo1}) and Eq.~(\ref{lo22}), we can write
\bqa
\Gamma^{(1)}_0&=&-{1\over f}{\partial V_0\over\partial\alpha}\;,
\label{eom11}
\eqa
where $\Gamma_n^{(1)}$ is the one-point function at order ${\cal O}(p^{2n+2})$.
We will next show that this relation is satisfied at next-to-leading
order, i.e. to order $n=1$.

\begin{figure}[htb]
\begin{center}\begin{tikzpicture}[line width=1.2 pt, scale=1.0]

	\draw[sigma] (3,0)--(4,0);
	\draw[sigma] (5,0) arc (360:0:0.5);

      \end{tikzpicture}
\caption{One-loop tadpole diagram contributing to the one-point function.}
      \label{tad}
    \end{center}
    \end{figure}
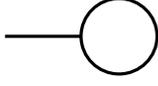

The one-loop diagrams contributing to the one-point function 
$\Gamma^{(1)}$
arise from the cubic terms in Eq.~(\ref{cubic}) and shown in Fig.~\ref{tad}.
All three pions run in the loop.
In order to work
consistently to next-to-leading order, 
the vertex factors must be
evaluated at the classical minimum,
${\partial V_0\over\partial\alpha}=0$ or
$\cos\alpha={m^2\over\mu_I^2}$.
After Wick rotation, we find
\bqa\nonumber
\Gamma_{\rm one-loop}^{(1)}|_{{\partial V_0\over\partial\alpha}=0}
&=&-{3m^2\sin\alpha\over2f}
\int_P{P^2+m_2^2\over(p_{0}^{2}+E_{\pi^+}^{2})
  (p_0^{2}+E_{\pi^-}^{2})}
\\ &&\nonumber
-{m^2\sin\alpha\over2f}
\int_P{P^2+m_1^2\over(p_{0}^{2}+E_{\pi^+}^{2})(p_{0}^{2}+E_{\pi^-}^{2})}
\\&&\nonumber
-{2\mu_I\sin\alpha\over f} m_{12}
\int_P{p_0^2\over(p_{0}^{2}+E_{\pi^+}^{2})(p_{0}^{2}+E_{\pi^-}^{2})}
\\ &&\nonumber
-{m^2\sin\alpha\over2f}\int_P{1\over P^2+m_3^2}
\\ \nonumber
&=&-\frac{1}{2f}\int_{P}\frac{1}{(p_0^{2}+E_{\pi+}^{2})(p_{0}^{2}+E_{\pi^{-}}^{2})}\\ \nonumber
&&\times\left [\frac{\partial m_{1}^{2}}{\partial\alpha}(P^{2}+m_{2}^{2})+\frac{\partial m_{2}^{2}}{\partial\alpha}(P^{2}+m_{1}^{2})
\right.\\ \nonumber&&\left.
  +\frac{\partial m_{12}^{2}}{\partial\alpha}p_{0}^{2}
\right ]
-\frac{m^{2}\sin\alpha}{2f}\int_{P}\frac{1}{P^{2}+m_{3}^{2}}
\\ 
&=&-{1\over f}{\partial V_1\over\partial\alpha}\;,
\label{onepoint}
\eqa
where $E_{i}=\sqrt{p^2+m_i^2}$ ($i=1,2$)
and the one-loop effective potential is
given by 
\bqa\nonumber
V_1&=&{1\over2}\int_P\log\left[(P^2+m_1^2)(P^2+m_2^2)+p_0^2m_{12}^2\right]
\\ &&
+{1\over2}\int_P\log\left[P^2+m_3^2\right]\;.
\eqa
Finally, comparing Eqs.~(\ref{lagstat}) and ~(\ref{laglinear}) it is
easy to see that
\bqa
{\Gamma^{(1)}_{\rm linear}}=-{\partial V_1^{\rm static}\over\partial\alpha}\;.
\label{linne}
\eqa
Adding Eqs.~(\ref{eom11}), (\ref{onepoint}), and (\ref{linne}),
we find
\bqa
\Gamma^{(1)}&=&-{1\over f}{\partial V_{\rm eff}\over\partial\alpha}=0\;,
\eqa
at the minimum of $V_{\rm eff}$. This is the unrenormalized version of
the equation of motion. We have checked that the divergences of the
one-loop diagram are cancelled by the counterterms up renormalization
of the couplings $l_i$.

\section{Expansion in $\alpha$}
\label{alpharekkje}
In this section, we consider the expansion of $V_{\rm eff}$ in powers
of $\alpha$. 
We begin with the tree-level term, which is
\bqa\nonumber
V_0&=&-f^2m^2+
  {1\over2}f^2\left(m^2-\mu_I^2\right)\alpha^2
  -{1\over24}f^2\left(m^2-4\mu_I^2\right)\alpha^4
  \\&&
+{\cal O}(\alpha^6)\;.
\label{alpha11}
\eqa
The static term $-{\cal L}_4^{\rm static}$ reads
\bqa\nonumber
V_{1}^{\rm static}&=&-(l_3+l_4)m^4
+\left[(l_3+l_4)m^4-l_4m^2\mu_I^2\right]\alpha^2
\\ &&\nonumber
-\left[{1\over3}(l_3+l_4)m^4
  -{5\over6}l_4m^2\mu_I^2+(l_1+l_2)
  \mu_I^4
\right]\alpha^4
\\ &&
+{\cal O}(\alpha^6)\;.
\label{alpha22}\eqa
Now consider
the NLO contribution from the charged pions
\bqa
V_{1,\pi^{\pm}}&=&
{1\over2}\int_P\log\left[(P^2+m_1^2)(P^2+m_2^2)+p_0^2m_{12}^2\right]\;,
\eqa
which can be rewritten as
\bqa\nonumber
V_{1,\pi^{\pm}}&=&
{1\over2}\int_P
\log\left\{\left[P^2+{1\over2}(m_1^2+m_2^2)\right]^2+p_0^2m_{12}^2
\right.\\ &&
\left.
  -{1\over4}(m_1^2-m_2^2)^2\right\}\;.
\eqa
Since $m_1^2-m_2^2=\mu_I^2\sin^2\alpha$, we proceed by expanding in powers of 
the mass difference, which as we will see is effectively the same as expanding
in powers $\alpha$. At $\mathcal{O}(\alpha^4)$, this yields
\bqa\nonumber
V_{1,\pi^{\pm}}&=&
{1\over2}\int_P\log\left
  \{\left[P^2+{1\over2}(m_1^2+m_2^2)\right]^2+p_0^2m_{12}^2\right\}
\\  \nonumber
&&-{1\over8}(m_1^2-m_2^2)^2\int_P
{1\over\left[P^2+{1\over2}(m_1^2+m_2^2)\right]^2+p_0^2m_{12}^2}\;.
\\ &&
\label{splittt}
\eqa
The first integral in Eq.~(\ref{splittt})
which we denote by $V_{1,\pi^{\pm}}^a$ can be performed by
rewriting the argument of the log in the integrand as
$[(p_0+{1\over2}{im_{12}})^2+p^2+{1\over2}(m_1^2+m_2^2+{1\over2}m_{12}^2)]
$\\$
\times
[(p_0-{1\over2}{im_{12}})^2+p^2+{1\over2}(m_1^2+m_2^2+{1\over2}m_{12}^2)]$.
Then by shifting the integration variable
$p_0\rightarrow p_0\mp {1\over2}{im_{12}}$
in the first and second pieces respectively, the integral can be written as
\bqa\nonumber
V_{1,\pi^{\pm}}^a&=&
\int_P\log
\left[P^2+{1\over2}\left(m_1^2+m_2^2+{1\over2}m_{12}^2\right)\right]
\\
&=&-{\tilde{m}^4\over2(4\pi)^2}\left[
{1\over\epsilon}+{3\over2}+\log{\Lambda^2\over\tilde{m}^2}
\right]\;,
\eqa
where
\bqa
\tilde{m}^2={1\over2}\left(m_1^2+m_2^2+{1\over2}m_{12}^2\right)=m^2\cos\alpha+{1\over2}\mu_I^2
\sin^2\alpha\;.
\eqa
Expanding to $\mathcal{O}(\alpha^{4})$ yields
\bqa\nonumber
V_{1,\pi^{\pm}}^a&=&
-{m^4\over2(4\pi)^2}\left[{1\over\epsilon}+{3\over2}
  +\log{\Lambda^2\over m^2}
\right]
\\ \nonumber
&&+{m^2(m^2-\mu_I^2)\over2(4\pi)^2}\left[{1\over\epsilon}+1
  +\log{\Lambda^2\over m^2}
\right]\alpha^2
\\ \nonumber
&&-{1\over24(4\pi)^2}\left[(4m^4-10m^2\mu_I^2+3\mu_I^4)
  \left({1\over\epsilon}+\log{\Lambda^2\over m^2}\right)
\right.\\ &&\left.
+m^4-4m^2\mu_I^2
\right]\alpha^4\;.
\label{alpha445}
  \eqa
The second integral (labelled as $V_{1,\pi^{\pm}}^b$) reads
\bqa\nonumber
V_{1,\pi^{\pm}}^b
&=&
-{1\over8}(m_1^2-m_2^2)^2
\int_P
{1\over\left[P^2+{1\over2}(m_1^2+m_2^2)\right]^2+p_0^2m_{12}^2}\;.
\\ &&
\eqa
Since the prefactor $(m_1^2-m_2^2)^2$ is $\mathcal{O}(\alpha^4)$ and higher,
the masses in the integral can be evaluated at $\alpha=0$
since we only care to expand the effective potential up to
$\mathcal{O}(\alpha^{4})$.
Using Eq.~(\ref{standard2}), we find
\bqa\nonumber
V^{b}_{1,\pi^{\pm}}&=&
-\frac{\mu_{I}^{4}}{8(4\pi)^{2}}\left[\frac{1}{\epsilon}+1
  \right.\\ \nonumber &&\left.
  -2\log\left (\frac{\sqrt{m^{2}-\mu_{I}^{2}}+\sqrt{m^{2}+\mu_{I}^{2}}}
      {\sqrt{2}\Lambda} \right ) 
\right.\\ &&\left.
    +{m^2\over\mu_I^2}-\frac{\sqrt{m^{2}-\mu_{I}^{2}}
      {\sqrt{m^{2}+\mu_{I}^{2}}}}{\mu_I^{2}} \right ]\alpha^4\;.
\eqa
The last contribution is given by Eq.~(\ref{lastcon}).
\bqa\nonumber
V_{1,\pi^0}&=&-{m^4\over4(4\pi)^2}\left[
{1\over\epsilon}+{3\over2}+\log{\Lambda^2\over m^2}
  \right]
\\ && \nonumber
  +{m^2(m^2-2\mu_I^2)\over4(4\pi)^2}\left[{1\over\epsilon}+1+\log{\Lambda^2\over m^2}\right]
  \alpha^2
  \\&& \nonumber
  -  {1\over48(4\pi)^2}\left[
    4\left(m^4-5m^2\mu_I^2+3\mu_I^4\right)
    \left({1\over\epsilon}+\log{\Lambda^2\over m^2}\right)
\right.\\ &&\left.
+m^4-8m^2\mu_I^2  \right]  \alpha^4\;.
  \label{alphazz}
  \eqa
 Adding Eqs.~(\ref{alpha11}),~(\ref{alpha22}), ~(\ref{alpha445}),
 and~(\ref{alphazz}), we can write the one-loop effective potential up
 to $\mathcal{O}(\alpha^{4})$:
\bqa\nonumber
  V_{\rm eff}^{\rm LG}&=&V_{\rm eff}(\alpha=0)+\frac{1}{2}m^{2}f^{2}
  \left [1-\frac{m^{2}}
    {2(4\pi)^{2}f^{2}}(\bar{l}_{3}-4\bar{l}_{4}) \right ]\alpha^{2}\\
  \nonumber
  &&-\frac{1}{2}f^{2}\mu_{I}^{2}\left [1+\frac{2m^{2}}{(4\pi)^{2}f^{2}}\bar{l}_{4}
  \right ]\alpha^{2}\\ \nonumber
  &&-\frac{1}{24}f^{2}\left [(m^{2}-4\mu_{I}^{2})-\frac{1}{2(4\pi)^{2}f^{2}}
    \left \{6\mu_{I}^{2}\sqrt{m^{4}-\mu_{I}^{4}}\right.\right.\\
\nonumber
  &&\left.\left.-10m^{2}\mu_{I}^{2}(3-4\bar{l}_{4})+4m^{4}\left(\frac{9}{4}+
        \bar{l}_{3}-4\bar{l}_{4}\right)\right.\right.\\
\nonumber
  &&+8\mu_{I}^{4}\left (\frac{9}{4}-\bar{l}_{1}-2\bar{l}_{2}\right.\\
  &&\left.\left.\left.+\frac{3}{2}\log\frac{\sqrt{m^{2}-\mu_{I}^{2}}+\sqrt{m^{2}+
          \mu_{I}^{2}}}{\sqrt{2}m} \right ) \right\} \right ]\alpha^{4}\;.
\label{readoff}
\eqa
In order to find the critical isospin chemical potential, we set the
coefficient of the $\mathcal{O}(\alpha^{2})$ term to zero and find that
$\mu_{I}=m_{\pi}$ at NLO, which can be found using
Eqs.~(\ref{mpi}).--(\ref{fpi}).
Then evaluating the $\mathcal{O}(\alpha^{4})$ term at this critical chemical
potential gives $a_{4}(\mu_{I}^{c})$ of Eq.~(\ref{LG2}), which is positive and
therefore at NLO the phase transition remains second order as at tree level.

\end{document}